\documentclass[conference]{IEEEtran}
\IEEEoverridecommandlockouts
\usepackage{cite}
\usepackage{amsmath,amssymb,amsfonts}
\usepackage{mathtools}
\usepackage{algorithmic}
\usepackage{graphicx}
\usepackage{textcomp}
\usepackage{caption}
\usepackage{subcaption}
\usepackage{xcolor}
\def\BibTeX{{\rm B\kern-.05em{\sc i\kern-.025em b}\kern-.08em
    T\kern-.1667em\lower.7ex\hbox{E}\kern-.125emX}}

\graphicspath{{./Figures/}}

\begin{document}

\title{State of Health Estimation of Lithium-Ion Batteries in Vehicle-to-Grid Applications Using Recurrent Neural Networks for Learning the Impact of Degradation Stress Factors}

\author{\IEEEauthorblockN{Kotub Uddin}
\IEEEauthorblockA{\textit{Sterling \& Wilson Renewable Energy,} \\
26 Brompton Square,\\
London SW3 2AD, UK.\\
kotub.uddin@sterlingwilson.com}
\and
\IEEEauthorblockN{James Schofield}
\IEEEauthorblockA{
\textit{Kaluza - An OVO company,}\\ 
140-142 Kensington Church Street,\\ London, W8 4BN, UK.\\
james@schofield.ai
\and
\IEEEauthorblockN{W. Dhammika Widanage}
\IEEEauthorblockA{\textit{WMG, University of Warwick,}\\
Coventry, CV4 7AL, UK. \\
Dhammika.Widanalage@warwick.ac.uk}
}}

\maketitle

\begin{abstract}
This work presents an effective state of health indicator to indicate lithium-ion battery degradation based on a long short-term memory (LSTM) recurrent neural network (RNN) coupled with a sliding-window. The developed LSTM RNN is able to capture the underlying long-term dependencies of degraded cell capacity on battery degradation stress factors. The learning performance was robust when there was sufficient training data, with an error of $<5$\% if more than 1.15 years’ worth of data was supplied for training.
\end{abstract}


\section{Introduction}
As transport and electrical-grid systems become more invested in electrification, the role of electrical energy storage is becoming more and more pronounced. Particularly lithium-ion batteries (LiBs); their relatively high energy and power densities coupled with their competitive cost and cycle life position them as the preferred commercial solution for electrical energy storage. Anticipating further ubiquity of lithium ion batteries, revenue streams are emerging where the flexibility of electrical storage \cite{b1,b2} is being commercialised. In application – given that beyond time ($t$) and number of cycles ($N$), LiB degradation is sensitive to temperature ($T$), state of charge (SoC), depth of discharge (DoD) and charge/discharge rate (C-rate) \cite{b3} – the gains generated through providing flexibility services can be offset by the cost of LiB degradation. Economic viability dictates therefore, that the revenue generated by providing flexibility must be (at the very least) larger than the incurred cost of battery degradation. To satisfy this online, state of health (SoH) estimation is required. 

    The difficulty with SoH predictions is path dependency \cite{b4}. That is, the dependency of SoH on the precise chronology of usage. Accounting for path dependency is not inherent to model based methods – whether they are based on fundamental electrochemical equations such as in \cite{b5} or circuit based equivalent models combined with filtering techniques such as in \cite{b6} – because path dependency is still not well understood.  Moreover, capturing long-term SoH dependencies on degradation stress-factors ($T$, SoC, DoD and C-rate) within such frameworks require a large amount of data to be collected over long periods of time, as demonstrated in \cite{b7}.
    
    A more natural approach is to employ data-driven methods that also employ a mechanism to store and update key information from the degradation data (as it becomes available online) via the effective learning of long-term dependencies. An example of this are Kernel methods, of which support vector machines (SVMs) is the most prominent. Such techniques have already been applied for SoH estimation \cite{b8,b9}. The limitation of Kernel methods is that in selecting the dominant data points to represent the degradation evolution of LiBs, prediction accuracy is lost (because path dependency is ‘diluted’) at the expense of computational efficiency. 
    
Neural Network (NN) based techniques, particularly recurrent neural networks (RNNs), are promising because they have an internal state that can represent path information.  Liu et. al. \cite{b10} demonstrated that the adaptive RNN showed better learning capability than classical training algorithms, including Kernel methods. The key impediment of classical RNNs however – common to all NNs with gradient-based learning methods and back propagation – is the vanishing gradient problem. That is, because the weights of the NN are updated proportional to the partial derivative of the error function with respect to the current weight in each iteration of training, the gradient will be vanishingly small, effectively preventing the weight from learning. The long short-term memory (LSTM) RNN, which is a deep learning NN, is designed to circumvent the vanishing gradient problem through the introduction of a \textit{forget gate} \cite{b11}. The LSTM RNN was applied and validated for SoH estimation by Zhang et al. \cite{b12}. Their work provides valuable detail and discussion on LSTM RNNs and the process of applying it for SoH estimation; interested readers are directed to \cite{b12} for more detail.

While the work of Zhang et al. \cite{b12} clearly demonstrates the strengths of LSTM RNNs, the authors chose not to allow the model to learn dependencies of SoH on LiB degradation stress factors. As demonstrated for home energy storage devices \cite{b13} and vehicle to grid applications \cite{b14}, optimising SoH with respect to $T$, SoC, DoD and C-rate is crucial to ensure the economic viability of flexibility as a revenue generating service. In this paper therefore, we extend the univariate approach adopted by Zhang et al. \cite{b12} to a multivariate time series that includes learning long-term dependencies of SoH on LiB degradation stress factors.

\section{Battery Degradation Data}
\begin{figure}
\centering
\includegraphics[width=0.5\textwidth]{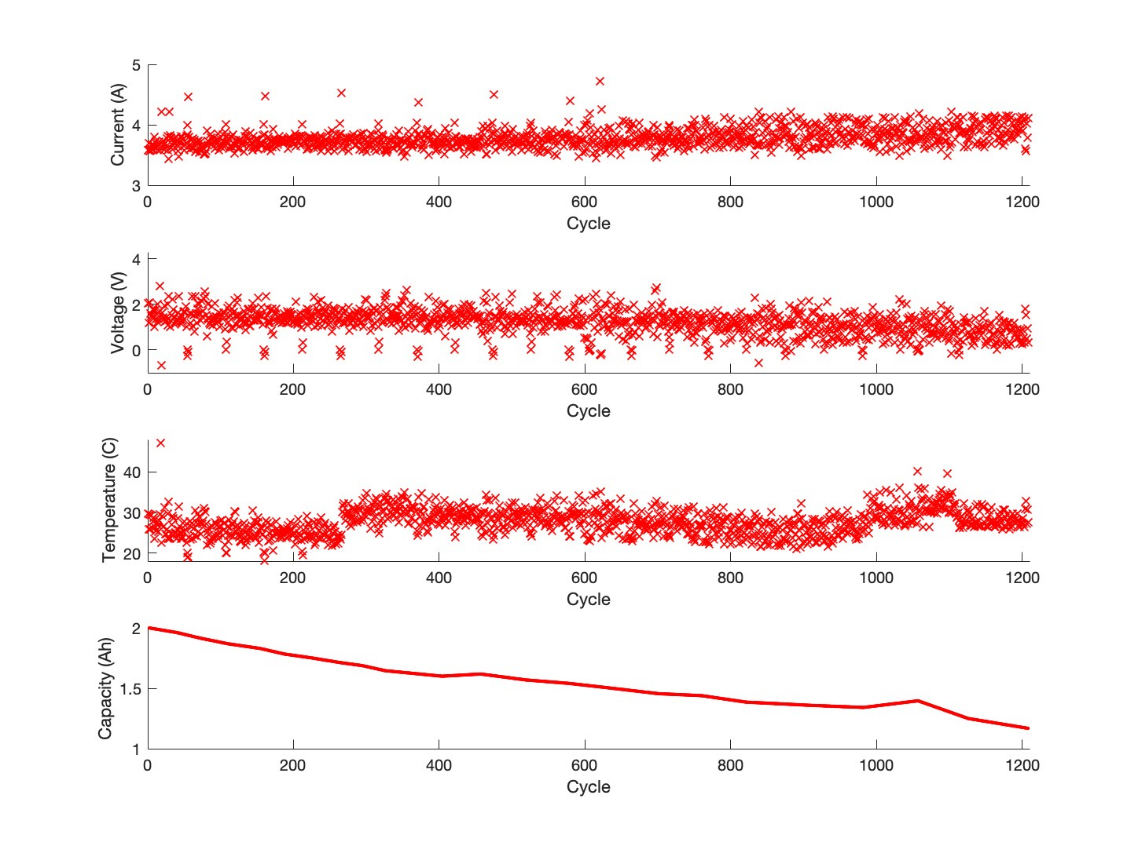}
\caption{Showing the average current, voltage and temperature over a cycle for the RW1 (in the numbering and labelling of the online repository) dataset. Capacity is the maximum discharge capacity measured at cycle $N$.}
\label{fig:RW1}
\end{figure}

The dataset used in this study is obtained from the NASA Ames Prognostics Center of Excellence Randomized Battery Usage Data Set \cite{b15}. The experiments consisted of applying randomized sequences of current loads ranging from 0.5 A to 4 A to a number of 18650 type cells produced by Idaho National Laboratory. The sequence is randomized in order to better represent real battery usage. After every 50 random walk steps the battery was characterised. The characterisation tests included a reference discharge at 1 C: the batteries were first charged with the constant current (2 A) and constant voltage (4.2 V) protocol. The test batteries were then discharged at 2 A until the battery voltage reaches 3.2 V. Pulsed current discharge, which consists of a 1 A load applied for 10 min followed by 20 min of rest, was also used to characterise resistance rise. Further detail of the testing procedure employed for this dataset can be found in \cite{b16}.

There were twenty-three sets of data available in this data repository, of which all except RW2, RW3 and RW18 (in the numbering and labelling of the online repository) were chosen for analysis. The datasets that were not used displayed anomalous temperature behaviour. Figure \ref{fig:RW1} displays the RW1 dataset post processing. Matlab was used for processing the raw data and to calculate the average current, voltage and temperature over each cycle to generate the required stress-factors for the LSTM RNN model. The stress-factors were number of cycles, average current, average voltage and average temperature, which were then mapped onto the corresponding cell capacity.

\section{LSTM RNN Architecture}
Literature describing the procedural and mathematical detail of LSTM RNNs is readily available online. For readability, in this section we summarise the most pertinent aspects of LSTM RNNs with a particular focus on highlighting any choices in architecture. In this work PyTorch \cite{b17}, an open-source machine learning library for Python, is used to implement the NN for time-series forecasting. PyTorch was preferred over Tensor Flow, Lasagne and Keras  \cite{b17} because: (i) PyTorch’s dynamic graph definition allows for variable length input sequences when implementing RNN architectures, (ii) the frameworks architectural style is relatively more native to Python and therefore allows for shorter development iterations, (iii) the default define-by-run mode is akin to traditional programming, (iv) common debugging tools such as pdb, ipdb or PyCharm debugger are supported, and (v) the platform supports declarative data parallelism. 

\subsection{Governing equations}
\begin{figure}
\centering
\includegraphics[width=0.4\textwidth]{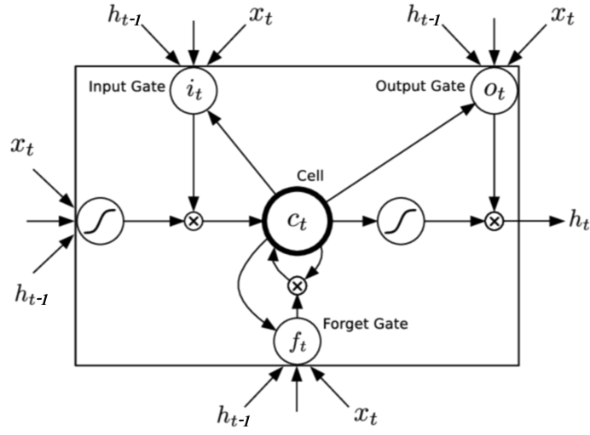}
\caption{Depicting an LSTM RNN cell}
\label{fig:LSTM}
\end{figure}

As mentioned previously, a simple RNN suffers from a fundamental problem of not being able to capture long-term dependencies in a sequence (i.e., the vanishing gradient problem). The LSTM framework resolves this through an internal state variable, which is passed from cell-to-cell and modified by operational gates, which are sigmoidal units activated through the current input layer $x_t$ as well as the hidden layer at the previous step $h_{t-1}$. In Figure \ref{fig:LSTM} the network architecture of an LSTM RNN predictor is schematically shown. 

The forget gate is defined by:
\begin{align}
f_t=\sigma\left(W_f^xx_t+W_f^hh_{t-1}+b_t\right).
\end{align}
The output of the forget gate is between 0 and 1. This output is multiplied with the internal state such that $f_t=0$ represents ‘forgetting’ the previous internal state completely while $f_t=1$ represents retaining the value completely. 

The input gate takes the previous output and the new input and passes them through another sigmoid layer. The input gate is defined by:
\begin{align}
i_t=\sigma\left(W_i^xx_t+W_i^hh_{t-1}+b_i\right)
\end{align}
where $W$ and $b$ values are the layer weights and biases of the forget and input gates, respectively. The input gate determines which values to update. The output $i_t$ is multiplied with the output of the candidate layer $c_t$:
\begin{align}
c_t=\tanh\left(W_c^xx_t+W_c^hh_{t-1}+b_c\right)
\end{align}
which returns a candidate vector to be added to the internal state $s_t$ such that:
\begin{align}
s_t={c_t}i_t+s_{t-1}f_t
\end{align}
Finally, the output gate, defined by:
\begin{align}
o_t&=\sigma\left(W_o^xx_t+W_o^hh_{t-1}+b_o\right)\\
h_t&={\tanh\left(s_t\right)}^\ast o_t
\end{align}

\subsection{Model Architecture}
In order to access the suite of standard linear and nonlinear machine learning algorithms it is useful to reframe a time series forecasting problem into a supervised learning problem, i.e., a task of learning functions that map inputs to outputs based on example input-output pairs \cite{b18}. This reframing, for both univariate and multivariate time series datasets, can be achieved through the introduction of a sliding window \cite{b19}. In this technique, a variable number, p, previous observations are considered as inputs of the network for each training process while the output is the forecasted, k, values of the time series. From a model predictive control (MPC) design perspective (which is used commonly for predictive controller design), the sliding window technique can be interpreted as a method that enables the use of SoH history to forecast future SoH trends up to a predefined prediction horizon. Assume that a given dataset represents the time-dependent profile of SoH from $t =0,1,dots,t_{end}$. SoH at time $t_n$ can be modelled using the SoH history over a finite set of previous observations, i.e. at times $t_{n-1},t_{n-2},dots,t_{n-p}$. It is worth pointing out that in implementation, the resulting algorithm is updated after each observation ($t’$) where $t’\in[t_{n+1},t_{n+z}]/t’\in\mathbb{R}$.

In place of classical stochastic gradient descent procedures employed for updating network weights iteratively, in this work we adopt the Adam optimization algorithm proposed by Kingma and Ba \cite{b20}. Adam combines  the advantages of two other extensions of stochastic gradient descent, namely: (i) The Adaptive Gradient Algorithm (AdaGrad) which maintains a per-parameter learning rate that improves performance on problems with sparse gradients (e.g. natural language and computer vision problems) and (ii) The Root Mean Square Propagation (RMSProp) that also maintains per-parameter learning rates that are adapted based on the average of recent magnitudes of the gradients for the weight. This means the algorithm does well on online and non-stationary problems (e.g. noisy).

Instead of adapting the parameter learning rates based on the average first moment (the mean) as in RMSProp, Adam also makes use of the average of the second moments of the gradients (the uncentered variance). Specifically, the algorithm calculates an exponential moving average of the gradient and the squared gradient, and the parameters $\beta_1$ and $\beta_2$ \cite{b20} control the decay rates of these moving averages. The initial value of the moving averages and $\beta_1$ and $\beta_2$ are chosen to be close to unity, which results in a bias of moment estimates towards zero. This bias is overcome by first calculating the biased estimates before calculating bias-corrected estimates.
Beyond the superior computational efficiency and little memory requirements, as demonstrated by Kingma and Ba \cite{b20} compared to its counterparts, Adam is well suited for problems that involve large datasets with noisy/or sparse gradients – which is appropriate for grid based applications. 

\section{Results and discussions}
\begin{figure}[h]
     \centering
     \begin{subfigure}[b]{0.3\textwidth}
         \centering
         \includegraphics[width=\textwidth]{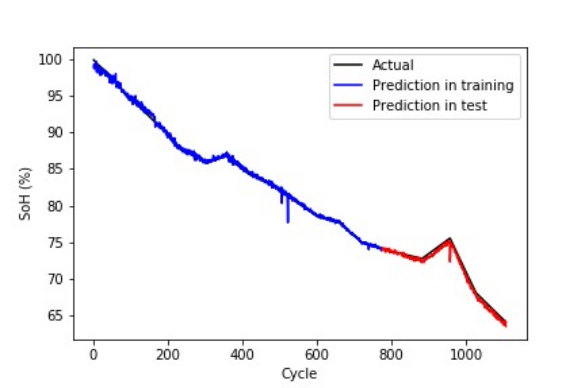}
         \caption{}
         \label{fig:TrainTest_a}
     \end{subfigure}
     \begin{subfigure}[b]{0.3\textwidth}
         \centering
         \includegraphics[width=\textwidth]{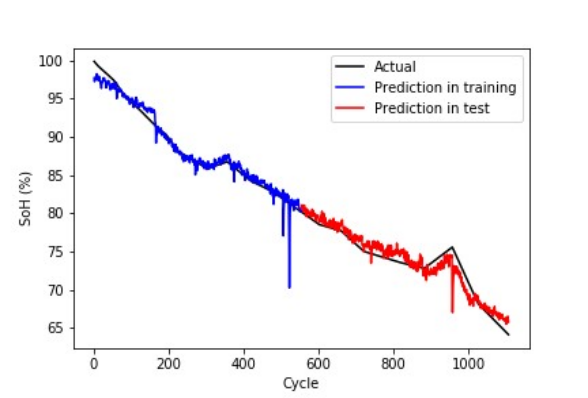}
         \caption{}
         \label{fig:TrainTest_b}
     \end{subfigure}
     \hfill
     \begin{subfigure}[b]{0.3\textwidth}
         \centering
         \includegraphics[width=\textwidth]{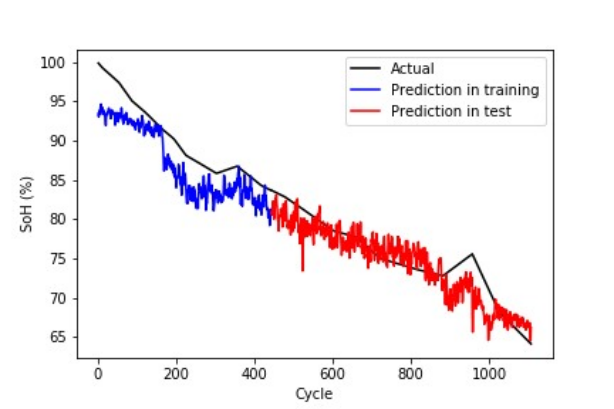}
         \caption{}
         \label{fig:TrainTest_c}
     \end{subfigure}
     \begin{subfigure}[b]{0.3\textwidth}
         \centering
         \includegraphics[width=\textwidth]{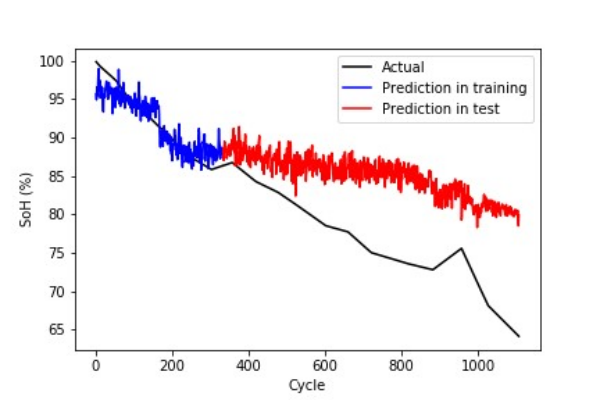}
         \caption{}
         \label{fig:TrainTest_d}
     \end{subfigure}
        \caption{Results for the LSTM RNN performance for training to testing ratio 7:3 (a), 5:5 (b), 4:6 (c) and 3:7 (d). }
        \label{fig:TrainTest}
\end{figure}

The datasets, obtained from the NASA Ames Prognostics Center, were split into ratios of 7:3, 1:1, 2:3 and 3:7 for training and testing, respectively. Results for the RW1 dataset (corresponding to Figure \ref{fig:RW1}) is shown in Figure \ref{fig:TrainTest}. The results show that, as expected, the accuracy of predictions in the test phase is highly sensitive to the length of the training data set, with a ratio of 7:3 exhibiting the best performance.  

    Assuming, in grid applications, that a cycle is completed daily – there is a requirement for at least 1.15 years’ worth of training data before SoH prediction accuracy is reasonable ($<5$\% error). This represents a long time, however, there is potential in the future to improve and reduce the learning time via transfer learning methodologies. For example, if a model was trained on the data from a fleet of vehicles with the same battery type, a new vehicle of that type could be initialised to the same parameters as the previously trained model.  
     
    In this work we use cycle number ($N$), cell current ($i$), cell terminal voltage ($V_{cell}$) and temperature as inputs. Although it’s typical to quote SoC as a degradation stress factor (usually because at rest, $V_{cell}$ is taken to be equivalent to open circuit voltage,  $V_{ocv}$ (SoC)), at an electrochemical level it is the electrode overpotentials ($\eta_{neg}$ and $\eta_{pos}$) that govern the growth of the solid electrolyte interphase, the dominant mode of LiB degradation \cite{b5,b21}. $V_{cell}$ is defined by  $\eta_{neg}$ and $\eta_{pos}$ and so it is appropriate to use $V_{cell}$ as an input. 
    It is worth noting that while open circuit voltage $V_{ocv}$ can be used as a proxy for SoC($V_{ocv}$), $V_{cell}$ cannot. This is because, even if we were to neglect diffusion and polarisation effects, the correspondence of $V_{cell}$ to  $V_{ocv}$ is convoluted by a nonlinear $iR(SoC,T)$ term (in a circuit analogy framework), where $R(SoC,T)$ is the cell’s Ohmic resistance. Given that the Bole et al.  \cite{b16} dataset includes slow rate discharge and charge data (which can be used as an approximate for $V_{ocv}$(SoC)) and current pulse data, an equivalent circuit model coupled with an associated filter can be developed to estimate SoC \cite{b22}. 

\section{Conclusions}
This work presents a lithium-ion cell SoH prediction tool that employs an LSTM RNN. The NN is trained and validated using NASA Ames Prognostics Center’s Randomized Battery Usage Data Set. The trained multivariate NN estimates LiB SoH by learning long term dependencies of LiB degradation stress factors and shows very good performance for a training and testing ratio of 7:3. The accuracy of test phase predictions are progressively worse for less training data. This poses a challenge because $>1.15$ years’ worth of data is needed for training to achieve the error target of $<5$\% (mean square error). 

    One possibility of addressing the inhibiting data requirements is to adopt a structured NN \cite{b23}. The internal parameter of the LSTM RNN lacks physical meaning in the hidden layers. A structured neural network (SNN) provides neurons with physical characteristics of the battery and allows access to internal states. Knowledge about the physical characteristics of a system can support the design of neural networks. As a result, it is possible to combine a high number of measured data/information and electrochemical knowledge or even an existing model of the system in an advantageous way. Since the neurons have physical meaning, it then becomes possible to initialize the weights with measured data or by expert knowledge in a way that the start values are already in the neighbourhood of the global minimum and local minima are avoided.
    
Another area of future work is to consider $\frac{dSoH}{dN}$ as the output as oppose to SoH. This will address the fluctuating nature of SoH presented in Figure \ref{fig:TrainTest}. This is also more intuitive because in time (or within a cycle) the stress factors act to accelerate battery degradation, i.e., $\mathrm{\Delta SoH}$. 

\section*{Acknowledgment}
Author WDW would like to thank the Faraday Institution [EP/S003053/1 grant number FIRG003] for part funding his research in relation to this work.

\end{document}